\documentclass[11pt]{article}      
\author{Kris Krogh \\
Neuroscience Research Institute\\ University of California, Santa
Barbara, CA 93106, USA\\ email: k\_krogh@lifesci.ucsb.edu}

\title
{Gravitation Without Curved Space-time}

\date{July 2, 2006}

\begin{document}

\maketitle

\begin{abstract}
\normalsize A quantum-mechanical theory of gravitation is presented,
where the motion of particles is based on the optics of de Broglie
waves. Here the large-scale geometry of the universe is inherently
flat, and its age is not constrained to $<$ 13 Gyr. While this
theory agrees with the standard experimental tests of Einstein's
general relativity, it predicts a different second-order deflection
of light, and measurement of the Lense-Thirring effect in the
upcoming NASA experiment \mbox{Gravity Probe B.}

\end{abstract}

\vfill\eject

\section{Introduction}

Modern physics has two different representations of gravitation:
in addition to the geometric one of Einstein's general relativity,
there is also the quantum-mechanical description.  According to
general relativity's weak equivalence principle, the motion of a
test particle in a gravitational field is independent of its mass.
However, in quantum mechanics, the motion depends intimately on
particle mass.  The mathematical structures of the two
representations ``seem utterly incompatible," in the words of
Francis Everitt.

Weinberg~\cite{sw} suggests that the prevailing geometric model of
gravitation ``has driven a wedge between general relativity and
the theory of elementary particles."  He also points out that this
approach is unnecessary:
\begin{quote}
Einstein and his successors have regarded the effects of a
gravitational field as producing a change in the geometry of space
and time.  At one time it was even hoped that the rest of physics
could be brought into a geometric formulation, but this hope has
met with disappointment, and the geometric interpretation of the
theory of gravitation has dwindled to a mere analogy, which
lingers in our language in terms like ``metric," ``affine
connection," and ``curvature," but is not otherwise very useful.
The important thing is to be able to make predictions about images
on the astronomers' photographic plates, frequencies of spectral
lines, and so on, and it simply doesn't matter whether we ascribe
these predictions to the physical effect of gravitational fields
on the motion of planets and photons or to a curvature of space
and time.
\end{quote}

It's often held that, beyond describing gravitation, curved
space-time explains it.  On this basis, Einstein's theory is taken
to be superior to others based solely on potentials.  But it does
not explain how mass-energy results in such curvature, so one
unknown is only replaced by another. And, despite heroic efforts
by Einstein and others, no geometric basis has been found for
electromagnetism.  We are left with inconsistent representations
of these phenomena.

Gravity and electromagnetism are more closely related in the
theory introduced here.  {\em It is assumed the effects of
gravitational potentials do not come indirectly, via space-time
curvature, but from their direct influence on quantum-mechanical
waves.}  Beyond its immediate compatibility with quantum
mechanics, the mathematical description of gravity obtained is
simpler and more precise than the present one.  As shown below,
this theory agrees equally well with the usual experimental tests
of general relativity.  Also, it makes new predictions for future
experiments.

The Hubble redshift has been taken by many as the ultimate
vindication of Einstein's general relativity. (See Misner, Thorne
and Wheeler~\cite{mtw0}.) In the associated ``standard" Big Bang
model, the redshift is attributed to a curved, expanding
space-time. That model is contradicted now by various
observations.

Those include measurements of the distribution of galaxies, which
reveal no discernible large-scale curvature~\cite{ls}.  According
to Linde~\cite{al}, the discrepancy is approximately {\em sixty\/}
orders of magnitude. (Toward a flatter geometry, hypotheses of
inflation, strange dark matter, the cosmological constant, and now
strange dark energy have been introduced {\em post hoc}. But there
remains no explanation why space-time is not curved by the quantum
zero-point energy of the vacuum.)

 Also, from the redshifts and distances
of nearby spiral galaxies estimated by Mould {\em et
al.}~\cite{jrm}, the ``standard" Big Bang model puts the maximum
age of the universe at 13 Gyr. (Direct measurement of the distance
to the galaxy NGC4258 by Herrnstein {\em et al.}~\cite{jrh} now
indicates that age needs to be revised downward~\cite{em}.)
Tsujimoto, Miamoto and Yoshi~\cite{tmy} find that substantially
less than the ages of stars in globular clusters.

This new theory leads instead to an evolutionary cosmology, in
which the Hubble redshift can be attributed to gradual change in
basic properties of the universe and atomic spectra.  This gives a
universe older than its stars, with an inherently flat geometry.

\section{Gravitational Potential}

At the most fundamental level, electromagnetism is described in
terms of the effects of potentials on the phases of
quantum-mechanical waves.  The phase shift $\Delta S$ of de
Broglie waves associated with a charged particle is given by

\begin{equation}
\Delta S \:=\: \frac{\,q}{\,h} \int \!\Phi \; dt -
\frac{\,q}{\,hc} \int \!{\bf A} \cdot \;d {\bf s}
\label{eq:1}
\end{equation}
where $\Delta S$ is measured in cycles, and the integrals are
taken over a possible trajectory, ${\bf s}$.  Using the Gaussian
system of units, $\Phi$ and ${\bf A}$ are the scalar and vector
electromagnetic potentials, $q$ is charge, $h$ Planck's constant,
and $c$ the speed of light.  As pointed out by Aharanov and
Bohm~\cite{ab} (and reiterated by Feynman~\cite{fls1}), this
subsumes the familiar Lorentz equation for the force on a charged
particle,
\begin{equation}
{\bf F}\:=\: q \left( {\bf E} + \frac{{\bf v} \times {\bf B}}{c}
\, \right)
\label{eq:2}
\end{equation}

Colella, Overhauser and Werner~\cite{cow}, demonstrated in 1975
that de Broglie waves are influenced similarly by gravitational
potentials.  That experiment measured the gravitational phase
shift of neutron waves.  The result was consistent with

\begin{equation}
\Delta S \:=\: \frac{\,m}{\,h} \int \!\Phi_{\!g} \; dt
\label{eq:3}
\end{equation}
where $m$ is the neutron mass and the subscript $g$ indicates a
gravitational potential.

Such effects are taken as the basis of gravitation here. While not
fully described by this equation, within experimental accuracy,
the same phase shift is predicted for the
Colella-Overhauser-Werner experiment. As in existing quantum
mechanics, the motion of particles will be found by applying
Huygens' principle to de Broglie waves, without introducing an
additional geodesic principle.

In rectangular coordinates, the scalar electromagnetic potential for
a particle at the origin, moving in the $x$ direction, is \pagebreak
\begin{equation} \Phi \:=\, \frac{q}{\sqrt{\,x^2 + (y^2
+z^2)(1-v^2/\,c^2)}} \label{eq:4}
\end{equation}
where $v$ is the velocity. And there is a vector potential,
\begin{equation}
{\bf A} \:=\, \frac {\,\bf v}{\,c} \,\Phi
\end{equation}

Unlike the two potentials in electromagnetism, or {\em ten} in
Einstein's gravity, there (currently) is only one gravitational
potential in this theory. We'll take it to have the same
relativistic form as the electromagnetic scalar. With the role
analogous to charge played by its inertial rest mass $m_0$, the
gravitational potential due to a small mass element is

\begin{equation}
\Phi_{\!g} \:=\, \frac{-\;G\,m_0}{\sqrt{\,x^2 + (y^2
+z^2)(1-v^2/\,c^2)}} \label{eq:6}
\end{equation}
where $G$ is the gravitational constant. (The inertial rest mass
includes the usual relativistic contributions of its moving
particles.) The equipotential surfaces then have the same shape and
arrangement for the three potentials $\Phi$, ${\bf A}$,
$\Phi_{\!g}$. (Consequently, they may be attributable to a unified
source~\cite{bk}.)

While these potentials themselves behave similarly and superpose in
the same linear fashion, the gravitational one will be seen to
differ in its nonlinear effects.  As described in Section 10, {\em
despite the absence of a gravitational vector potential, there are
velocity-dependent effects on moving bodies.} Those sometimes
resemble magnetism. In the case of the lunar orbit, there is an
equivalent of the gravitomagnetic effect in Einstein's theory, as
shown in Section 11.

Electromagnetic potentials are governed by the wave equations
\begin{equation}
\nabla^2 \Phi \,-\, \frac{1}{c^2}\frac{\partial^2\Phi}{\partial
t^2} \:=\: -\,4 \pi \rho
\end{equation}
and
\begin{equation}
\nabla^2 {\bf A} \,-\, \frac{1}{c^2}\frac{\partial^2{\bf
A}}{\partial t^2} \:=\: -\,\frac{4 \pi \bf j}{c}
\end{equation}
To satisfy special relativity, the gravitational potential is taken
to obey an equation of the same form.  From the corresponding
quantities in Eqs.~(\ref{eq:4})\ and\ (\ref{eq:6}), that gives
\begin{equation}
\nabla^2 \Phi_{\!g} \,-\,
\frac{1}{c^2}\frac{\partial^2\Phi_{\!g}}{\partial t^2} \:=\: 4 \pi
G \rho_m  \label{eq:9}
\end{equation}
where $\rho_m$ refers to the density of the mass appearing in
Eq.~(\ref{eq:6}).

The gravitational waves this describes have the same velocity $c$ as
in general relativity. For a single gravitational potential, they
are simple longitudinal waves like those in acoustics. Because $c$
will depend on $\Phi_{\!g}$, they are also nonlinear. As in the
acoustic wave approximation, the waves can be treated as linear for
small amplitudes such as those found in the solar system.

In his search for the basis of general relativity's geometry,
``pre-geometry," Wheeler has called attention to a proposal by
Sakharov~\cite{ads} that gravitation ultimately arises from
variation in the quantum zero-point energy of the vacuum. (See
Puthoff~\cite{hep1}.) Sakharov's conjecture is that space-time
curvature is determined by the distribution of vacuum energy. We'll
adopt a similar hypothesis: that gravitational potentials correspond
to regions of diminished vacuum energy, which determines the
velocity of quantum-mechanical waves.  Like the speed of sound in a
gas, the velocity is less where the energy density is
lower~\cite{bk}.

\section{Wave Velocities and Transformations}

In 1911, before arriving at the current theory of general
relativity, Einstein~\cite{ae1} proposed that the effect of a
gravitational potential is a decrease in the speed of light. While
accurately describing the gravitational redshift and the behavior
of clocks, that theory predicted only half the measured value for
the deflection of starlight by the Sun.

In this theory, a reduction of the velocities of {\em all\/}
quantum-mechanical waves, including light, is taken as the
fundamental effect of gravitational potentials. The amount of
reduction, and the way this is manifested differ substantially
from Einstein's initial scheme.  Here, the speed of light is given
by
\begin{equation}
c \:=\: c_0 e^{2 \Phi_{\!g} / c_0^2} \label{eq:10}
\end{equation}
where $c_0$ is the value in the absence of a gravitational
potential.

As shown in Appendix A, this function can be derived directly from
general principles. One is an extended principle of relativity,
holding that the observed laws of physics are also unchanged for
reference frames in uniform gravitational potentials. (However,
there is no assumption of the equivalence principle, in any of its
various forms. Nor is there any assumption of a geodesic principle,
general covariance, or Mach's principle.) Eq.~(\ref{eq:10}) has also
been derived by Puthoff~\cite{hep2} from a model of the vacuum.

From the extended principle of relativity, the velocity $V$ of de
Broglie waves slows in proportion to light in a gravitational
potential. Hence we can also write

\begin{equation}
V \:=\: V_0 e^{2 \Phi_{\!g} / c_0^2} \label{eq:11}
\end{equation}
where the $0$ subscript again indicates the corresponding quantity
without a potential.  While these wave velocities are defined with
respect to a preferred reference frame (such as that identified by
the cosmic microwave background), the system nevertheless obeys
special relativity where the gravitational potential is uniform. The
special relativity adopted here is that advocated by Lorentz and
Poincar\'{e}, and more recently by Bell~\cite{jsb}.

(Bell saw this preferred-frame relativity as a likely prerequisite
for a causal quantum mechanics. An advantage is its provision of a
preferred direction for time and entropy. For that purpose,
parameterized quantum mechanics introduces a time parameter $\tau$
into Minkowski space-time, as a function of a Newtonian time $t$.
Hartle~\cite{jbh} has noted it may be impossible to incorporate such
a parameter into general relativity. However, in
Lorentz-Poincar\'{e} relativity, $t = \tau$ and no added parameter
is needed.)

At the quantum-mechanical level, the frequencies of de Broglie waves
determine the rates of clocks, and their wavelengths the sizes of
atoms and measuring rods.  In special relativity, the frequencies of
constituent particles in moving atoms, and their wavelengths in the
direction of motion, diminish by the same factor.  Here a similar
effect exists in gravitational potentials.

The de Broglie wavelength $\lambda$ and frequency $\nu$ are related
to wave velocity by
\begin{equation}
V \:=\: \lambda \,\nu \label{eq:12}
\end{equation}
As derived in Appendix A, $\lambda$ and $\nu$ diminish identically
as
\begin{equation}
\nu \:=\: \nu_0 e^{\Phi_{\!g}/c_0^2} \label{eq:13}
\end{equation} and
\begin{equation}
\lambda \:=\: \lambda_0 e^{\Phi_{\!g}/c_0^2} \label{eq:14}
\end{equation} where the subscript $0$ has the same
meaning as before.  (In this case, the change in $\lambda$ is
isotropic, and doesn't occur in just one dimension.)

It follows the rates of clocks and lengths of measuring rods change
in gravitational potentials by the factor $e^{\Phi_{\!g}/c_0^2}$.
Consequently, the locally measured speed of light remains constant.
These represent changes in {\em objects themselves} rather than the
geometry of space-time.

``Relativity" was Poincar\'{e}'s term, and to illustrate the
principle~\cite{hp} he asked: What if you went to bed one night, and
when you awoke the next day everything in the world was a thousand
times bigger?  Would you notice anything?  He pointed out such
effects aren't observed locally, since measuring devices change with
the objects they measure.  As Einstein noted~\cite{ae4}, he also
insisted the true geometry of the universe is Euclidean.  Objects
behave here as in a Poincar\'{e} world.

We can also derive transformations for energy and mass.  In accord
with the extended principle of relativity, Planck's constant $h$ is
taken to be invariant in a gravitational potential. From the
relation $E = h \nu$ and Eq.~(\ref{eq:13}), a particle's rest energy
$E_0$ varies with its de Broglie frequency as
\begin{equation}
E_0 \:=\: E_{00} e^{\Phi_{\!g}/c_0^2} \label{eq:15}
\end{equation}
where $E_{00}$ represents the energy for both a zero velocity and
zero gravitational potential.  Since its energy is the sum of its
particles', this also holds for a macroscopic body.

From Einstein's relation, the last equation can be expressed as
\begin{equation}
m_0 c^2  \:=\: m_{00} c_0^2 e^{ \Phi_{\!g}/c_0^2} \label{eq:16}
\end{equation}
where similarly $m_0$ and $m_{00}$ are the rest mass and rest mass
without a gravitational potential.  Substituting for $c$ from
Eq.~(\ref{eq:10}) gives the transformation of rest mass in a
gravitational potential,
\begin{equation}
m_0 \:=\: m_{00} e^{-3 \Phi_{\!g}/c_0^2} \label{eq:17}
\end{equation}
This is also the mass responsible for gravitational potentials
appearing in Eq.~(\ref{eq:6}).

\section{Optics of Light and Matter Waves}

To calculate trajectories, the basic method we'll use dates
(remarkably) to Johann Bernoulli (1667-1748). Bernoulli discovered
the motion of a body in a gravitational field can be treated as an
optics problem, through assuming a fictitious refractive index,
proportional to the square root of the difference between its
kinetic and potential energies~\cite{cl1}.  The same deep analogy
between mechanics and optics underlies the Hamiltonian
representation of classical mechanics.  Hamilton based that on
hypothetical surfaces of constant ``action" orthogonal to the
trajectories of bodies.

The ``optical-mechanical analogy" also had a central role in the
development of wave mechanics by de Broglie and Schr\"{o}dinger
~\cite{tlh}. There the analogy becomes more direct: instead of fixed
surfaces of constant action, there are moving de Broglie wavefronts.
Like light rays, the trajectories of particles are orthogonal to
those. Consequently, when diffraction and interference effects can
be neglected, the trajectories of particles or bodies can be found
by the methods of geometrical optics.

To arrive at matter waves, de Broglie began by treating massive
particles as Planck oscillators~\cite{gl}. Equating the Planck and
Einstein energies for a particle
\begin{equation}
E \:=\: h \nu \:=\: m c^2 \label{eq:18}
\end{equation}
its frequency $\nu$ is $m c^2/h$.  Next, from the Lorentz transform
for a moving oscillator, de Broglie showed the velocity $V$ for
waves matching a moving particle's phase is
\begin{equation}
V \:=\: \frac{\,c^2}{\,v} \label{eq:19}
\end{equation}
where $v$ is the particle's velocity.  Its wavelength $\lambda$ then
is $V/\nu$ or
\begin{equation}
\lambda \:=\: \frac{\,h}{\,mv}
\end{equation}

Note the relationship between $V$ and $v$ in Eq.~(\ref{eq:19})
doesn't depend on mass. In this gravity theory, it obviates the
usual assumption of the weak equivalence principle, which dictates
that a test particle's motion is mass independent.

Interference in beams of atoms and molecules shows these equations
hold beyond elementary particles.  In principle, as emphasized by
Wheeler~\cite{wf}, large bodies also have de Broglie wavelengths --
although from the last equation they become too short to observe. In
that case diffraction and interference are negligible, and
geometrical optics can be used to describe the waves.

The variational principle in geometrical optics is Fermat's
principle of least time.  From Eq.~(\ref{eq:19}), matter particles
and their waves have different velocities.  What's minimized is the
travel time for {\em wavefronts} between two points, with respect to
other trajectories.  (At a more fundamental level, Fermat's can be
viewed as a consequence of Huygens' principle. Unlike general
relativity, we'll introduce no additional variational principle
beyond those already in quantum mechanics.)

Fermat's principle can be expressed as
\begin{equation}
\delta \int_{s_0}^s n \: ds \:=\: 0
\end{equation}
where $s$ refers to distance over some path connecting two fixed
points and $n$ is the refractive index (inverse of wave velocity) as
a function of $s$.

As Marchand~\cite{ewm} shows, for any spherically symmetric index
gradient, the resulting path in polar coordinates is
\begin{equation} \theta \:=\: \theta_0 +
k\int_{r_0}^r \frac{dr}{r\,\sqrt{r^2 n^2 - \,k^2 }} \label{eq:22}
\end{equation}
where $\theta_0$ and $r_0$ are the values at an arbitrary starting
point. The quantity $k$ is a constant given by
\begin{equation}
k \:=\: \pm r n \sin \psi \label{eq:23}
\end{equation}
where, for any point on the ray trajectory, $\psi$ is the angle
between the trajectory and a radial line connecting the point and
origin.

The last two equations can be used to describe both light rays and
the trajectories of bodies in central gravitational fields.  We'll
introduce the notation
\begin{equation}
\mu \:=\: \frac{GM}{c_0^2} \label{eq:24}
\end{equation}
where $G$ is the gravitational constant and $M$ the mass of a
gravitational source.  For a spherical source body at rest, the
dimensionless potential becomes
\begin{equation}
\frac{\Phi_{\!g}}{c_0^2} \:=\: -\frac{GM}{c_0^2\,r} \:=\:
-\frac{\:\mu}{\:r} \label{eq:25}
\end{equation}

The speed of light from Eq.~(\ref{eq:10}) is then
\begin{equation}
c \:=\: c_0 e^{-2 \mu /r} \label{eq:26}
\end{equation}
giving the spherically symmetric refractive index
\begin{equation}
n \:=\: \frac{\,c_0}{\,c} \:=\: e^{2\mu /r} \label{eq:27}
\end{equation}
Putting this expression for $n$ into Eq.~(\ref{eq:22}) gives the
path of a light ray near a massive body.

As done by de Broglie for charged particles~\cite{ldb2}, we can
construct a similar index of refraction for matter waves
\begin{equation}
n \:=\: \frac{\,c_0}{\,V} \label{eq:28}
\end{equation}
(Again, this is a dimensionless number inversely proportional to the
wave velocity. Only its relative variation matters, so its scale is
arbitrary.) We'll use it here to find the trajectory of a test body
in a central gravitational field.  For that we'll need an expression
for $V$ as a function of the body's radial position.

To put $V$ in terms of energy, we can use the relativistic transform
\begin{equation}
E \:=\:  \frac{E_0}{\sqrt{1\,-\,v^2\!/c^2}} \label{eq:29}
\end{equation}
where $E_0$ again is the body's rest energy.  Solving for $v$ gives
\begin{equation}
v \:=\:  c \,\sqrt{1\,-E_0^2/E^2} \label{eq:30}
\end{equation}
Inserting this into Eq.~(\ref{eq:19}), we get
\begin{equation}
V \:=\:  \frac{c}{\sqrt{1\,-\,E_0^2/E^2}} \label{eq:31}
\end{equation}

While the principle of energy conservation says $E$ is constant for
a freely orbiting body, $E_0$ depends on the gravitational
potential.  From Eqs.~(\ref{eq:15}) and~(\ref{eq:25}), \mbox{$E_0$
is}
\begin{equation}
E_0 \:=\:  E_{00}\,e^{-\mu/r} \label{eq:32}
\end{equation}
And from this and Eq.~(\ref{eq:26}) for $c$, we can rewrite the
previous equation as
\begin{equation}
V \:=\:  \frac{c_0 e^{-2\mu/r}}
{\sqrt{1\,-\,\frac{E_{00}^2}{E^2}\,e^{-2\mu/r}}}
\end{equation}
in terms of orbital constants and the single variable $r$.
Eq.~(\ref{eq:28}) becomes
\begin{equation}
n \:=\:\sqrt{\:e^{4\mu/r}\,-\,\frac{E_{00}^2}{E^2}\:e^{2\mu/r}}
\end{equation}

Putting this $n$ into Eq.~(\ref{eq:22}) gives an exact equation for
the orbital trajectory
\begin{equation}
\theta \:=\: k\int \frac{dr}{r\, \sqrt{\,r^2 \left( e^{4\mu/r}
\,-\:\frac{\,E_{00}^2}{\,E^2}\;e^{2\mu/r} \right) - \,k^2}}
\label{eq:35}
\end{equation}
(As the velocity of a particle or body approaches $c$, the quantity
$E_{00}^2/E^2$ vanishes and this becomes the light ray trajectory
found previously.)  Again, from the derivation of Eq.~(\ref{eq:22}),
$k$ is a constant.  Here it plays essentially the same role as
conserved angular momentum in Newtonian mechanics.

To describe $k$, we can express $n$ as a function of the orbital
velocity $v$. Rewriting Eq.~(\ref{eq:19}) in terms of $c_0$ via
Eq.~(\ref{eq:26}), then substituting the resulting expression for
$V$ into Eq.~(\ref{eq:28}) we get

\begin{equation}
n \:=\: \frac{v e^{4\mu/r}}{c_0}
\end{equation}
Inserting this $n$ into Eq.~(\ref{eq:23}), $k$ is given by
\begin{equation}
k \:=\: \frac{r v e^{4\mu/r} \sin \psi}{c_0} \label{eq:37}
\end{equation}
where $\psi$ is the angle between the orbital velocity vector and
radial position vector. \pagebreak

To illustrate the observer problem in quantum mechanics, both
Einstein and Wheeler have asked whether the Moon is there when no
one looks. Regardless of the interpretation of quantum mechanics
adopted, its usual rules apply here to large bodies.  That includes
de Broglie's requirement that orbital lengths correspond to integral
numbers of wavelengths. For a body following the Moon's approximate
orbit, the possible states are effectively infinite in number and
indistinguishable. But {\em in principle} they're discrete, like an
atomic electron's.

As in quantum optics, Huygens' principle could be used instead of
Fermat's for calculating wave amplitudes and the probability a body
goes from point A to point B in a gravitational field.  Section 9
outlines how such amplitudes can be calculated using the basic
methods of quantum electrodynamics, from the phases of bodies over
various possible paths. (Preferably small bodies like neutrons.)

\section{Predicted Frequency Shift}

For metric theories of gravity, the PPN formalism of Will and
Nordtvedt~\cite{mtw4} provides a convenient way of checking them
against an array of experimental tests, in terms of an isotropic
space-time. Metric theories have come to be defined as any obeying
the equivalence principle.  However, even under that broad
definition, this isn't a metric theory and the extent to which the
PPN formalism can be applied isn't clear.

The alternative pursued in this paper is to compare the theory's
predictions directly against each existing test of general
relativity.  While we'll use isotropic coordinates, in this case no
conversion from curved to isotropic space-time is needed. (Of course
measurements still need to be corrected for the condition of the
measuring devices, to match those of observers removed from
gravitational potentials.)

Although preferred-frame effects are possible in this theory, it can
be shown they are small for existing Solar System experiments.  For
simplicity, this paper treats the Solar and Earth-Moon systems as
being at rest in the preferred frame of the universe. (Corrections
for their additional motions will be given in a subsequent paper.)

Where $\nu$ represents the rate of a clock near a massive spherical
body at rest, from Eqs.~(\ref{eq:13}) and~(\ref{eq:25}), the series
expansion for the exponential gives
\begin{equation}
\nu \:=\: \nu_0 e^{-\mu/r} \:=\: \nu_0 \!\left( 1 - \frac{GM}{c_0^2
r}+ \frac {1}{\,2}\! \left( \frac{GM}{c_0^2 r} \right)^2 + \cdot
\cdot \cdot \,\right) \label{eq:38}
\end{equation}

Unlike this theory, gravitational potentials in general relativity
don't vary simply as $\!1/r$ in isotropic coordinates. (And are not
analogous to electromagnetic potentials in this respect.) The rates
of clocks also differ from Eq.~(\ref{eq:13}) as functions of those
potentials. {\em However, these differences effectively cancel, and
in terms of the isotropic radius, the clock rates predicted by
Einstein's final theory are the same to the second
order}~\cite{mtw2}. Both theories agree well with the best direct
measurement of the gravitational frequency shift to date, by NASA's
Gravity Probe A~\cite{rv}, which is only accurate to the first order
in $GM/(c^2 r)$.

\section{Precession of Mercury's Orbit}

Mercury's orbit is given by Eq.~(\ref{eq:35}). To solve that for the
orbital precession,  we can take just the first three terms of the
series expansions for $e^{4\mu/r}$ and $e^{2\mu/r}$, since higher
powers of $\mu/r$ are vanishing in the Solar System. The result is
\begin{equation}
\theta \:=\: k\int \frac{dr}{r\, \sqrt{\,r^2 \left(
 1+\frac{4\mu}{r}+\frac{8\mu^2}{r^2}-\frac{E_{00}^2}{E^2}
 -\frac{2\mu E_{00}^2}{r E^2}-\frac{2\mu^2 E_{00}^2}{r^2 E^2}
 \right) - \,k^2 }}
\end{equation}
Multiplying terms by $r^2$ inside the square root gives a quadratic
there,
\begin{equation}
\theta \:=\: k \int \frac{dr}{r\,\sqrt{
\left(1-\frac{E_{00}^2}{E^2}\right)r^2
+\left(4\mu-2\mu\frac{E_{00}^2}{E^2}\right)r
+\left(8\mu^2-2\mu\frac{E_{00}^2}{E^2}-k^2\right)}}
\end{equation}

We'll use these notations for the quadratic coefficients:
\begin{equation}
A \:=\: 1 -\, E_{00}^2/E^2 \label{eq:41}
\end{equation}

\begin{equation}
B \:=\: 4\mu \,-\: 2\mu E_{00}^2/E^2
\end{equation}

\begin{equation}
C \:=\: 8\mu^2 -\: 2\mu^2 E_{00}^2/E^2-\:k^2 \label{eq:43}
\end{equation}
From these and a table of integrals, we get
\begin{eqnarray}
\theta & = & k \int \frac{dr}{r \sqrt{Ar^2+Br +C}}\nonumber\\
 & = & \frac{k}{\sqrt{-\,C}} \:\sin^{-1}\!\left(\frac{Br+2C}{r
\sqrt{B^2-\,4AC}}\right) \label{eq:44}
\end{eqnarray}

Solving for r gives the time-independent equation for the orbit:
\begin{equation}
r \:=\: \frac{-2\,C/B}{\,\,1-\,\sqrt{1-\,4AC/B^2}\;\sin \left(
\theta \sqrt{-\,C/k^2}\right)} \label{eq:45}
\end{equation}
This has the basic form of a polar equation of an ellipse,
\begin{equation}
r \:=\: \frac{a\,(1-\,\epsilon^2\,)}{1-\,\epsilon \,\sin \theta}
\end{equation}
where $a$ is the semi-major axis and $\epsilon$ is the eccentricity.
(Eq.~(\ref{eq:45}) describes hyperbolic trajectories also.) In
Eq.~(\ref{eq:45}), the quantity corresponding to $a(1-\epsilon^2)$
(called the {\em parameter}, or $p$, in orbital mechanics) is given
by
\begin{eqnarray}
a\,(1-\,\epsilon^2\,) & = & -2\,C/B\nonumber\\
& = & \frac{-2 \left(8\mu^2-\,2\mu^2 E_{00}^2/E^2-\,k^2 \right)}
 {4\mu-\,2\mu E_{00}^2/E^2}
\end{eqnarray}

Since Mercury's  orbital velocity is effectively nonrelativistic,
$E_{00}^2/E^2$ is extremely close to one.  Also, the value of
$\mu^2$ is minuscule compared to $k^2$.  In this case, the last
equation reduces to
\begin{equation}
a\,(1-\,\epsilon^2\,) \:=\: k^2/\,\mu \label{eq:48}
\end{equation}
Calling the total change in $\theta$ from one minimum of r
(perihelion) to the next $\Delta \theta$, it follows from
Eq.~(\ref{eq:45}) that
\begin{equation}
\Delta\theta\sqrt{-\,C/k^2} \:=\: 2\pi
\end{equation}
Again, since the value of $E_{00}^2/E^2$ is very nearly one,
Eq.~(\ref{eq:43}) becomes
\begin{equation}
C \:=\: 6\mu^2-k^2
\end{equation}
Rearranging and making this substitution for $C$ gives
\begin{equation}
\Delta\theta \:=\: 2\pi /\sqrt{1-\,6\mu^2/k^2}
\end{equation}
We can take just the first two terms of the binomial expansion for
the inverse square root, since higher powers of $\mu^2/k^2$ are
vanishing.  This gives
\begin{equation} \Delta\theta \:=\: 2\pi
\!\left(1+3\mu^2/k^2 \right)
\end{equation}
Then substituting for $k^2$ via Eq.~(\ref{eq:48}), we arrive at
\begin{equation}
\Delta\theta \:=\: 2\pi+ \frac{6\pi\mu}{a\,(1-\,\epsilon^2\,)}
\label{eq:53}
\end{equation}

From the last term, the perihelion is shifted in the direction of
orbital motion.  This corresponds to Einstein's equation for the
orbital precession~\cite{ae2}, and agrees closely with the 43$''$
per century value observed for Mercury.

\section{Deflection of Light}

The gravitational frequency shift in Einstein's 1911
variable-speed-of-light theory~\cite{ae1} was
\begin{equation}
\nu \:=\: \nu_0\! \left( 1 + \frac{\Phi_{\!g}}{c^2} \right)
\end{equation}
which agrees with Eq.~(\ref{eq:13}) to the first order.  But there
was no effect on $\lambda$, or the dimensions of measuring rods,
corresponding to Eq.~(\ref{eq:14}). Consequently, the speed of light
in a gravitational potential was
\begin{equation}
c \:=\: c_0\! \left( 1 + \frac{\Phi_{\!g}}{c^2} \right)
\end{equation}

Einstein used a wave approach in his 1911 paper to derive the
deflection of starlight by the Sun and Jupiter.  From Huygens'
principle and this expression for $c$, he obtained the first-order
approximation
\begin{equation}
\alpha \:=\: \frac{2GM}{c^2 r}
\end{equation}
where $\alpha$ is the angular deflection in radians and $r$ is the
radial distance to the light ray at its closest point.

(This deflection agreed with Einstein's original equivalence
principle, which held that the observed laws of physics in a uniform
gravitational {\em field} are the same as those in an extended,
accelerating reference frame.  Later, he restricted this to
infinitesimal regions of space, discarding the spatial framework.
And in terms of isotropic coordinates, the gravitational bending of
light is no longer equivalent to that in an accelerating reference
frame.)

In this theory, from the series expansion for $c$
\begin{equation}
c \:=\: c_0\! \left( 1 + \frac{2\Phi_{\!g}}{c_0^2 }+
\frac{2\Phi_{g}^2}{c_0^4} + \cdot \cdot \cdot \right) \label{eq:57}
\end{equation}
Omitting the terms above first-order and using the same method, this
c gives a deflection twice as large,
\begin{equation}
\alpha_1 \:=\: \frac{4GM}{c_0^2 r} \label{eq:58}
\end{equation}
where $\alpha_1$ represents the first-order deflection in radians.
This is the first-order effect predicted by general relativity,
about $1.''75$ for a star near the Sun's limb.

Eqs.~(\ref{eq:22}) and~(\ref{eq:27}) give the exact path of a light
ray near a massive body. Illustrating the close analogy between
light and matter, the second-order deflection of light is given by
the solution of Eq.~(\ref{eq:35}) obtained for Mercury's orbit. For
the constant $k$, we set $r$ in Eq.~(\ref{eq:37}) equal to the light
ray's closest radial distance. We also set $\sin\psi$ to 1 and the
particle or body's velocity $v$ to the speed of light given by
Eq.~(\ref{eq:26}).

With that $k$, Eqs.~(\ref{eq:41})-(\ref{eq:44}) give the hyperbolic
ray trajectory in polar coordinates.  (For light, the terms
involving $E_{00}^2/E^2$ disappear.)   Setting $r$ to infinity in
Eq.~(\ref{eq:44}), $\theta$ is then the deflection for half the ray
trajectory.  The resulting first-order deflection for the complete
trajectory again is that in Eq.~(\ref{eq:58}). Taking the sign of
$\alpha_1$ as positive, the second-order term $\alpha_2$ can be
expressed as
\begin{equation}
\alpha_2 \:=\: - \frac{1}{\,2} \,\alpha_1^2
\end{equation}
when $\alpha_1$ and $\alpha_2$ are in radians.  (Since $\Phi_g$ is
negative, the same relationship between first and second-order terms
is seen in Eq.~(\ref{eq:57}).) For a star near the Sun's limb,
$\alpha_2$ represents a decrease of about 7.4 $\mu$arcseconds.

In contrast, general relativity predicts a second-order decrease of
about $3.5$ $\mu$arcseconds~\cite{rm}.  It also predicts an
additional deflection due to the Sun's rotation and dragging of
space-time~\cite{ghz}. That would depend on the Sun's angular
momentum, which is uncertain, but the effect would be at least
$+0.7$ and $-0.7$ $\mu$arcseconds on opposite sides of the Sun. Such
an effect is absent in this theory.

The proposed NASA/ESA experiment LATOR (Laser Astrometric Test Of
Relativity)~\cite{sgt} would measure the Sun's gravitational
deflection of light to high accuracy.  Using triangular laser
ranging, between two probes on the far side of the Sun and a laser
transponder on the International Space Station, the expected
accuracy is 1 part in $10^3$ of general relativity's second-order
light deflection -- far more than needed to choose between these
theories.

\section{The Lunar Orbit}

In many alternative theories of gravity, such as the Brans-Dicke
theory, the \mbox{predicted} accelerations of bodies depend on their
masses and binding energies. Nordtvedt pointed out in 1968 that if
the accelerations of the Earth and Moon toward the Sun differed,
there would be an additional oscillation in the Moon's orbit which
could be measured by lunar laser ranging~\cite{kn}.  The oscillation
amplitude exceeds a meter in some cases, and such theories have been
ruled out by the range measurements, which are accurate now to the
order of a few centimeters.

Einstein's theory also predicts a small difference in the
accelerations, corresponding to a 3 cm oscillation~\cite{bs},
which is consistent with the lunar ranging data.  This is due to
the Moon's orbital motion around Earth, and without that, the
post-Newtonian approximation of Einstein's theory predicts
identical solar accelerations for the Earth and Moon in isotropic
coordinates. That would be the case if the two bodies orbited the
Sun separately at the same radius, and we will compare this
theory's predictions for that example next.

To include the effects of a body's own mass in the equations of
motion, we introduce a new variable $P$.  This will represent the
potential due to its mass, divided by $c_0^2$ to make it
dimensionless.  We take $P$ to describe the average potential seen
by the body's various mass elements, treating it as a whole.

To find the Sun's effect on $P$, we express its value of
$\Phi_{\!g}/c_0^2$ again as $-\mu/r$. Putting that into
Eq.~(\ref{eq:17}), the rest mass of a nearby body is increased by
the factor $e^{3\mu/r}$. Also, from Eq.~(\ref{eq:14}), the body's
radius decreases by $e^{-\mu/r}$. Since the potential contributed by
each of its own mass elements is given by $-\,G m_0/r$, the rest
case for $P$ is described by
\begin{equation}
P_0 \:=\:  P_{00} e^{4\mu/r} \label{eq:60}
\end{equation}
where $P_{00}$ is the value when the body's velocity and Sun's
potential are both zero.  Then from the relativistic
transformation of a scalar potential,
\begin{equation}
P \:=\:  \frac{P_0}{\sqrt{1\,-\,v^2\!/c^2}} \:=\: \frac{P_{00}
e^{4\mu/r}}{\sqrt{1\,-\,v^2\!/c^2}} \label{eq:61}
\end{equation}

With the value of $\Phi_{\!g}/c_0^2$ given now by $P-\mu/r$, we'll
determine the velocity and acceleration of a body following Earth's
orbit.  The orbit has little eccentricity, and as done by Nordtvedt,
we will treat its shape as circular. As in Section 6, we can use
conservation of energy to find the velocity, although the
relationship between the total and rest energies will be slightly
different.

In this case, $E_{00}$ will represent the body's rest energy in the
absence of the Sun's potential, where its own potential is still
present.  To put the general rest energy $E_0$ in terms of $E_{00}$,
we need to account for the difference in its own binding potential,
in addition to that from the Sun.  Eq.~(\ref{eq:32}) then becomes
\begin{equation}
E_0 \:=\: E_{00} e^{(P_0-P_{00}-\mu/r)}
\end{equation}
Also, with the contributions of both the Sun and orbiting body to
the total potential, Eq.~(\ref{eq:26}) for the speed of light
becomes
\begin{equation}
c \:=\: c_0 e^{2(P-\mu/r)}
\end{equation}
Substituting these expressions for $E_0$ and $c$ into
Eq.~(\ref{eq:30}) gives the velocity
\begin{equation}
v \:=\:  c_0 \,e^{2\,(P\,-\,\mu/r)}
\left(1\,-\,\frac{\,E_{00}^2}{\,E^2}\:e^{2\,(P_0\,-\,P_{00\,}-\,\mu/r)}\right)^{1/2}
\label{eq:64}
\end{equation}

From de Broglie's $V=c^2/v$, the corresponding wave velocity is
\begin{equation}
V \:=\:  c_0 \,e^{2\,(P\,-\,\mu/r)}
\left(1\,-\,\frac{\,E_{00}^2}{\,E^2}\:e^{2\,(P_0\,-\,P_{00\,}-\,\mu/r)}\right)^{-1/2}
\end{equation}
Again, the wavefronts are orthogonal to a trajectory, and are
arranged radially around the Sun for a circular orbit. V increases
exactly in proportion to r, and the condition for a circular orbit
is
\begin{equation}
\frac{\:d}{\,dr} \left(\!\frac{\:V}{\:r}\right) \:=\:  0
\end{equation}
Taking the derivative, solving for $E_{00}^2/E^2$, then substituting
for $E_{00}^2/E^2$ in Eq.~(\ref{eq:64}), the resulting velocity is
\begin{equation}
v \:=\:  c_0 \,e^{2\,(P\,-\,\mu/r)} \left( {\frac{\mu/r +\, r\,(d
P_0 / d r)  } {1 -\, \mu/r \,+\, r\,(d P_0/ d r) -\, 2\,r\,(d P/ d
r)}}\right)^{1/2} \label{eq:67}
\end{equation}

We will need an estimate for $P$, which depends in part on $v$. The
result of the above equation is strongly influenced by the term
$r\,(d P_0 / d r)$ in the numerator and the precise variation of
$P_0$, but this is not the case for $P$.  And since $v$ for the
Earth or Moon is almost the same as the velocity of a body with a
negligible gravitational potential, we can use that to estimate $P$.
In that case, Eq.~(\ref{eq:67}) gives
\begin{equation}
v \:=\:  c_0 \,e^{-2\,\mu/r}  \left( {\frac{\;\mu/r } {\,1 -\,
\mu/r}}\right)^{1/2} \label{eq:68}
\end{equation}
Also, $c = c_0 e^{-2\mu/r}$.  Replacing $v$ and $c$ in
Eq.~(\ref{eq:61}), we find
\begin{equation}
P \:\cong\: P_{00} \,e^{4 \mu/ r}  \left( {\frac{\,1-\mu/r } {\,1
-\, 2\,\mu/r}}\right)^{1/2} \:\cong\: P_{00} \,e^{9 \mu/(2 r) }
\end{equation}
Hence
\begin{equation}
\frac{\,d P}{\,dr} \:\cong\:  - \frac{\,9\,\mu\, P_{00} }{2\,
r^2}\;e^{9 \mu/(2 r)}
\end{equation}
From Eq.~(\ref{eq:60}), the exact derivative of $P_0$ is
\begin{equation}
\frac{\,d P_0}{\,dr} \:=\:  - \frac{\,4\,\mu\, P_{00}
}{\;r^2}\;e^{4 \mu/r}
\end{equation}

Putting these expressions for $P$, $d P/ d r$, and $d P_0/ d r$ into
Eq.~(\ref{eq:67}) gives $v$ entirely in terms of $c_0$, $\mu$, $r$
and $P_{00}$. Since these equations are already in isotropic
coordinates and the orbit is circular, the acceleration then is just
$a=v^2/r$.  $P_{00}$ is approximately twice the ratio of a body's
gravitational binding energy to its total relativistic energy. Using
$-9.\,2 \times 10^{-10}$ and $-0.\,4 \times 10^{-10}$ for the Earth
and Moon respectively, these values of $P_{00}$ give effectively
identical accelerations at Earth's orbital radius, differing by only
$1.\,9$ parts in $10^{17}$.

From the changed speed of light alone, appearing as the term $c_0
\,e^{2 (P\,-\,\mu/r)}$ in Eq.~(\ref{eq:67}), $P$ would introduce an
acceleration difference more than {\em eight orders of magnitude
greater}. But that is canceled almost perfectly by the effect of
change in the binding potential. (A similar cancelation occurs in
the post-Newtonian approximation of Einstein's theory.) The residual
difference in the accelerations translates~\cite{mn} to an
unobservable 0.6 $\mu$m oscillation in the lunar orbit. An
observable perturbation of the orbit is described in Section 11.

\section{Lagrangian and Quantum Mechanics}

In the tradition of metric theories of gravity, we could have begun
by simply writing the Lagrangians for matter and light, and using
those for calculations.  But we also want to show their physical
basis and the connection to quantum mechanics.

Compared to general relativity, the gravitational Lagrangian in this
theory is more closely related to that in electromagnetism.
Referring to Eq.~(\ref{eq:1}), substituting ${\bf v}dt$ for $d{\bf
s}$, the fundamental equation of electromagnetism can be rewritten
as
\begin{equation}
\Delta S \:=\: \frac{\,q}{\,h} \,\int (\,\Phi  -\, {\bf v} \cdot
{\bf A}\,/\,c\,) \:dt \label{eq:72}
\end{equation}
Again, over a possible path of a charged particle, this integral
gives the cumulative shift in the de Broglie phase due to
electromagnetic potentials.  As described by Feymnann~\cite{fls1},
summing the resulting phase amplitudes over all possible paths from
point A to point B gives the total amplitude and probability of
finding it at B.

We'll show that effect corresponds to the Lagrangian for a charged
particle
\begin{equation}
L \:=\, \,-\,m_0 c^2{\sqrt{1\,-\,v^2\!/c^2}} \:- \:q\,(\,\Phi -\,
{\bf v} \cdot {\bf A}\,/\,c\,) \label{eq:73}
\end{equation}
Note this equation lacks the property of manifest covariance
advocated by Einstein. (Einstein took the opposite position during
one period~\cite{ae3}, writing that the mathematical representation
of gravity ``cannot possibly be {\em generally} covariant.")
Nevertheless, the expression is properly relativistic, and as
Goldstein~\cite{hg} has noted, no manifestly covariant formalism has
been found for electrodynamics.

Again, de Broglie waves match a particle's phase at its moving
position.  Where $v$ is the particle's velocity, its frequency there
is
\begin{equation}
\nu_v \:=\; \nu\,'{\sqrt{1\,-\,v^2\!/c^2}} \label{eq:74}
\end{equation}
with $\nu\,'$ the de Broglie frequency in the inertial frame where
the particle is at rest.  Using the Planck and Einstein relations,
and the relativistic transformation for the potential, the
particle's energy in the primed frame can be expressed as
\begin{eqnarray}
h\nu\,' & = & m_0 c_0^2 \,+\, q\Phi'\nonumber\\
 & = & m_0 c_0^2 \,+\, q(\,\Phi  -\, {\bf v} \cdot
{\bf A}\,/\,c\,)\,/\sqrt{1\,-\,v^2\!/c^2}
\end{eqnarray}
Dividing both sides by $h$, then substituting for $\nu\,'$ in the
previous equation, we get
\begin{equation}
\nu_v \:=\; \frac{\:m_0 c^2}{\,h}\:{\sqrt{1\,-\,v^2\!/c^2}} \,+
\frac{\:q}{\:h}\:(\,\Phi  -\, {\bf v} \cdot {\bf A}\,/\,c\,)
\label{eq:76}
\end{equation}
corresponding to the phase shift $\Delta S$ in Eq.~(\ref{eq:72}).

Differing only by the constant factor $-h$, this and the Lagrangian
in Eq.~(\ref{eq:73}) represent {\em the same quantity}, expressed in
frequency or energy terms. Like the integral of the Lagrangian
(action), the variation of the de Broglie wave phase $S$ near a
particle trajectory is zero.  (Corresponding to orthogonal
wavefronts.)  As noted by Lanczos~\cite{cl1}, either gives the same
trajectory between two fixed points in space.

From Eq.~(\ref{eq:13}), in a gravitational potential, a body's de
Broglie frequency at its moving position becomes
\begin{equation}
\nu_v \:=\; \left(\frac{\:m_{00} c_0^2}{\,h}
\:{\sqrt{1\,-\,v^2\!/c^2}} \,+ \frac{\:q_0}{\:h}\:(\,\Phi_0  -\,
{\bf v} \cdot {\bf
A}_0\,/\,c\,)\right)e^{\Phi_{\!g}/c_0^2}\label{eq:77}
\end{equation}
where $q_0$, $\Phi_0$ and ${\bf A}_0$ are the values of $q$, $\Phi$
and ${\bf A}$ for a zero gravitational potential.  Without
electromagnetic potentials, the corresponding gravitational phase
shift is
\begin{equation}
\Delta S \:=\: \frac{\,m_{00} c_0^2}{\,h} \,\int
\sqrt{1\,-\,v^2\!/c^2} \,\left(\,e^{\Phi_{\!g}/c_0^2} \,-\, 1
\right)\:dt \label{eq:78}
\end{equation}
which can be used to calculate probability amplitudes.  {\em For
weak fields and slow velocities, this reduces to Eq.~(\ref{eq:3}),
which describes the neutron interference experiment of Colella,
Overhauser and Werner~\cite{cow}.}

Multiplying the frequency of Eq.~(\ref{eq:77}) by $-h$ gives the
Lagrangian for this theory
\begin{equation}
L \,= \left(-\,m_{00} c_0^2\,{\sqrt{1\,-\,v^2\!/c^2}} \:-
\:q_0\,(\,\Phi_0 -\, {\bf v} \cdot {\bf A}_0\,/\,c\,)\right)
e^{\Phi_{\!g}/c_0^2} \label{eq:79}
\end{equation}
In the absence of electromagnetic potentials, this is the same
Lagrangian found by Puthoff~\cite{hep2}.  Here the full Lagrangian
has the property that, for any system of particles in a uniform
gravitational potential, all particles experience the same relative
effects. And special relativity is recovered in the limit where
gravitational fields go to zero, as in general relativity.

The Euler-Lagrange equations of motion are given, as in
electrodynamics, by
\begin{equation}
\frac {\partial L}{ \partial q_i} \,-\, \frac{\, d}{ d t} \left (
 \frac {\partial L}{\, \partial \dot q_i}\, \right) =\: 0
\end{equation}
where (rather than charge) $q_i$ represents a generalized coordinate
in isotropic space.  Appendix B shows these give the same orbital
equations obtained previously from de Broglie wave optics.

\section{Acceleration Equation}

With no electromagnetic potentials, the gravitational Lagrangian can
be written as
\begin{equation}
L \:=\: -\,E_{00}\: e^{\Phi_{\!g}/c_0^2} \,\sqrt{1-\left(\,v^2
/\,c_0^2\,\right)e^{-4\Phi_{\!g}/c_0^2}} \label{eq:81}
\end{equation}
expressing the term $m_{00} c_0^2$ as $E_{00}$, and substituting
for $c$ from Eq.~(\ref{eq:10}). As usual, the canonical momentum
${\bf p}$ is the partial derivative with respect to velocity (see
Duffey~\cite{ghd})
\begin{equation}
{\bf p} \:=\: \frac {\,\partial L}{\,\partial {\bf v}}\:=\: \frac
{\,{\bf v}\,E_{00}\, e^{-3\Phi_{\!g}/c_0^2}} {\,c_0^2
\,\sqrt{1-\left(\,v^2 /\,c_0^2\,\right)e^{-4\Phi_{\!g}/c_0^2}}}
\end{equation}

From Eqs.~(\ref{eq:29}) and (\ref{eq:32}),
\begin{equation}
E_{00} \:=\: E\,e^{-\Phi_{\!g}/c_0^2}\,\sqrt{1-\left(\,v^2
/\,c_0^2\,\right)e^{-4\Phi_{\!g}/c_0^2}} \label{eq:83}
\end{equation}
Putting this into the preceding equation gives
\begin{equation}
{\bf p} \:=\: \frac{{\,\bf v}\,E \,
e^{-4\Phi_{\!g}/c_0^2}}{\,c_0^2}
\end{equation} \label{eq:84}

$E$ is a constant for conservative systems, and the time
derivative of ${\bf p}$ is
\begin{equation}
\frac {d{\bf p}}{dt} \:=\:
\frac{\,E\,e^{-4\Phi_{\!g}/c_0^2}}{\,c_0^2}\left(\frac {\,d{\bf
v}}{\,dt} \,-\, \frac {\,4{\bf v} \,d{\Phi_{\!g}}}{\,c_0^2
\,dt}\right)
\end{equation}
To find the gradient of a Lagrangian, $v$ is held constant. Taking
the gradient of Eq.~(\ref{eq:81}) and then substituting for $E_{00}$
from Eq.~(\ref{eq:83}) we get
\begin{equation}
\nabla L \:=\: \frac{\,E\,e^{-4\Phi_{\!g}/c_0^2}}{\,c_0^2}\left(
-\nabla\Phi_{\!g}\,e^{4\Phi_{\!g}/c_0^2} - \frac
{\,v^2}{\,c_0^2}\,\nabla\Phi_{\!g} \right)
\end{equation}

From the relation $d{\bf p}/dt = \nabla L$~\cite{ghd}, we can equate
the right sides of the last two equations, eliminating the conserved
$E$. Then solving for $d{\bf v}/dt$ gives
\begin{equation}
{\bf a} \:=\: -\nabla\Phi_{\!g} \left(e^{4\Phi_{\!g}/c_0^2}+ \frac
{\,v^2}{\,c_0^2}\right) + \frac{\,4{\bf v}}{\,c_0^2} \left(\frac
{\,d{\Phi_{\!g}}}{\,dt}\right) \label{eq:87}
\end{equation}
where $d{\bf v}/dt$ is written as the acceleration $\bf a$.  This is
the gravitational counterpart of the Lorentz force equation in
electromagnetism, describing the relativistic acceleration of bodies
in gravitational fields.  (Multiplying both sides by the same mass
would give the force.) Note there are effects here that depend on a
body's relative velocity, as in electromagnetism.

To describe many body systems in general relativity, it's customary
to use a many-body Lagrangian~\cite{kn5}. Here, this acceleration
equation can be used instead. {\em Like the Lorentz force equation,
it can be employed to describe systems of any type, including those
that don't conserve energy.}

Its easily shown the equation agrees with Eq.~(\ref{eq:68}) for a
body in a circular orbit. We have ${\bf a} = -v^2/r$,
$\nabla\Phi_{\!g} = c_0^2\,\mu/r^2$, $\Phi_{\!g}/c_0^2 = -\mu/r\,$
and $\,d \Phi_{\!g}/dt = 0$. Eq.~(\ref{eq:87}) becomes
\begin{equation}
-\frac{\,v^2}{\,r} \:=\: -\frac{\,c_0^2\,\mu}{\,r^2}\left(
e^{-4\mu/r} + \frac {\,v^2}{\,c_0^2} \right)
\end{equation}
and solving for $v$ gives Eq.~(\ref{eq:68}).  It can also be shown
shown to agree with the general equations of motion for orbiting
test bodies given in Appendix B.

\section{Gravitomagnetism}

Shahid-Saless~\cite{bs} has shown that, for a Fermi (geocentric)
reference frame, general relativity predicts a gravitomagnetic
interaction between the Moon and Sun. Such an effect appears to
have been detected the lunar ranging experiment, and we'll
investigate this theory's prediction for that experiment next.

The ranging experiment gauges the lunar distance by the
time-of-flight of laser pulses originating from an Earth
observatory and returning from one of four reflectors on the Moon
(positioned by Apollo astronauts and the Russian Lunakhod II
spacecraft). It provides a comparison of the distances at the new
and full Moons. We'll estimate the predicted distances at those
points in the lunar orbit.  To compare our estimates with the
ranging experiment, we'll also need to account for the effect of
the Sun's gravity on $c$ and a laser pulse's time-of-flight.

Newtonian mechanics provides a remarkably accurate description of
the lunar orbit, and here we want to describe lunar distances with
respect to their corresponding Newtonian values.  Like
Shahid-Saless, we'll use a geocentric inertial frame, where the
Sun is in relative motion.  As done by Nordtvedt, we'll treat the
lunar orbit as lying in the plane of the ecliptic.

For simplicity, we'll also take the Moon's velocity component in the
Sun's direction to be zero at the new and full Moons. The term
$d\Phi_{\!g}/dt$ in Eq.~(\ref{eq:87}) vanishes and, for those points
in the lunar orbit, the Moon's acceleration due to the Sun's
potential alone can be written as
\begin{equation}
a_s \:=\: \frac{G M}{R^2} \left( e^{-4\, G M/(c_0^2 R)} + \frac
{\,v^2}{\,c_0^2}\right)
\end{equation}
where $M$ is the solar mass, $R$ is the Sun-Moon (not Sun-Earth)
distance, $v$ is the lunar velocity, and acceleration toward the
Sun is defined as positive. (This uses the slow-motion
approximation for the Sun's potential, which will not affect our
results.)

At the full Moon, the gradient of Earth's potential is in the same
direction as the Sun's, and when the effects of Earth's potential
are included we obtain
\begin{equation}
a_{s+e} \:=\: \left( \frac{G M}{R^2}+ \frac{G\, m}{r^2}\right)
\left( e^{- \frac{4\, G M}{c_0^2 R}\,- \frac{4\, G\, m}{c_0^2 r}}
+ \frac {\,v^2}{\,c_0^2}\right)
\end{equation}
where $m$ is Earth's mass, and $r$ is the Earth-Moon distance.
After expanding the exponentials in the last two equations, the
ratio of the lunar accelerations with and without Earth's
potential can be expressed accurately as
\begin{equation}
\frac{a_{s+e}}{a_s} \:\cong\: \frac {\frac{G M}{R^2}+ \frac{G\,
m}{r^2}}{\frac{G M}{R^2}} \left( 1- \frac{4\, G\, m}{c_0^2\,
r}\right) \label{eq:91}
\end{equation}

This equation differs from the corresponding Newtonian ratio only in
the last term.  We'll assume that, for a given $R$, when this ratio
of accelerations equals the Newtonian value, the observed curvature
of the lunar orbit is the same.  Then, where $r_0$ is the Newtonian
Earth-Moon distance, we ask what value of $r$ gives the same curved
trajectory relative to the Sun's position.  We'll write
\begin{equation}
\frac{G M}{R^2}+ \frac{G\, m}{r_0^2} \:=\: \left( \frac{G M}{R^2}+
\frac{G\, m}{r^2}\right) \left( 1- \frac{4\, G\, m}{c_0^2\,
r}\right) \label{eq:92}
\end{equation}
where we have put the Newtonian acceleration ratio on the left side,
Eq.~(\ref{eq:91}) on the right, and multiplied both sides by $G
M/R^2$. In the small last term of this equation, $r_0$ can be
substituted for $r$, since they are nearly identical. Then solving
for $r$ gives
\begin{equation}
r \:\cong\: r_0 \,-\, \frac{2\, G M r_0^2}{c_0^2\, R^2} \,-\,
\frac{2\, G\, m}{c_0^2} \label{eq:93}
\end{equation}

At the new Moon, the gradient of Earth's potential has the opposite
sign and Eq.~(\ref{eq:92}) becomes
\begin{equation}
\frac{G M}{R^2}- \frac{G\, m}{r_0^2} \:=\: \left( \frac{G M}{R^2}-
\frac{G\, m}{r^2}\right) \left( 1- \frac{4\, G\, m}{c_0^2\,
r}\right)
\end{equation}
where the solution for $r$ is now
\begin{equation}
r \:\cong\: r_0 \,+\, \frac{2\, G M r_0^2}{c_0^2\, R^2} \,-\,
\frac{2\, G\, m}{c_0^2} \label{eq:95}
\end{equation}

Alternatively, the same results can be obtained using the lunar
acceleration with and without the Suns's potential, for a given
lunar trajectory around Earth. Eq.~(\ref{eq:91}) becomes
$a_{s+e}\,/\,a_e$, the Newtonian Sun-Moon distance $R_0$ and $r$ are
given, and $R$ is solved for. Translating the difference found in
$R$ to one in $r$ having an equal effect on the lunar acceleration,
Eqs.~(\ref{eq:93}) and (\ref{eq:95}) are arrived at again.

In Eq.~(\ref{eq:93}) for the full Moon, we'll take $r_0$ equal to
the mean Earth-Moon distance, and $R$ to the mean Sun-Earth distance
plus $r_0$. Keeping the terms separate, the radius of the lunar
orbit in centimeters is then
\begin{equation}
r \:\cong\: \ r_0 - 1.9 - 0.9
\end{equation}
In Eq.~(\ref{eq:95}) for the new Moon, $R$ becomes the mean
Sun-Earth distance minus $r_0$, and we get
\begin{equation}
r \:\cong\: \ r_0 + 2.0 - 0.9
\end{equation}
From the middle terms, the Moon is 1.9 cm closer to Earth than its
Newtonian distance when full, and 2.0 cm more remote at the new
Moon, corresponding to a shift (polarization) of the lunar orbit
of about 2 cm toward the Sun. (The final $-0.9$ cm terms only
decrease the size of the orbit.)

The ranging experiment also sees an {\em apparent} shift in the
lunar orbit, due to variation in the speed of light over the paths
of the reflecting laser pulses.  At the full Moon's location, the
Sun's potential is weaker by approximately $G M r /R^2$ than at
Earth's position, and a pulse's propagation speed is increased, in
accord with Eq.~(\ref{eq:10}). Averaging the relative speed of
light over the beam path and translating that to a decrease in the
apparent lunar distance, the result is
\begin{equation}
\Delta r_a \:\cong\: -\frac{\,G M r^2}{\,c_0^2 R^2} \:\cong\:-1.0
\end{equation}
The effect at the new Moon is the opposite, giving a 1.0 cm
increase in the apparent distance, and the equivalent of an
additional 1 cm shift of the lunar orbit toward the Sun.

For Einstein's theory, Nortvedt~\cite{kn2} obtains an identical
``effective range perturbation" due to the Sun's influence on light
propagation. As he notes, in terms of isotropic coordinates, the
actual effects of that influence are changes in the pulse
propagation times. The delay of a laser pulse reflected at the new
Moon represents the well-known ``Shapiro effect," and is
indistinguishable here from that in general relativity.

Combining the actual and apparent perturbations of the Newtonian
lunar orbit, Shahid-Saless~\cite{bs} and Nortvedt~\cite{kn3}
obtain a total shift of 3 cm for Einstein's theory. The same
result is obtained here. Using almost identical sets of ranging
data, Williams, Newhall and Dickey~\cite{wnd} and M\"{u}ller and
Nordtvedt~\cite{mn} find that agrees with the observed orbital
shift to -0.8 $\pm$ 1.3 cm and +1.1 $\pm$ 1.1 cm respectively.

For gravitomagnetism affecting the Moon's trajectory, the
significant motions of the interacting bodies are effectively in a
single plane. That isn't so for a satellite in a polar orbit near
the rotating Earth, and there the predictions of Einstein's theory
and this one diverge.

According to Einstein, gravitational potentials don't act directly
on particles or their de Broglie waves, but on the space-time they
reside in.  In addition to the curvature caused by mass-energy,
gravitomagnetism is assumed to result from the dragging of
space-time by mass-energy currents. As first described by Lense
and Thirring~\cite{lt}, this would involve a local rotation of
space-time near a rotating, massive body.

In the low-speed, weak-field limit of general relativity, the effect
of space-time rotation on a test body's trajectory can also be
described in terms of a gravitational vector potential and a
corresponding gravitomagnetic field~\cite{cw}.  For an Earth
satellite in a polar orbit, the field induces a varying acceleration
perpendicular to the satellite's orbital plane, and a precession of
that plane in the direction of Earth's rotation.

This theory (currently) has no gravitational vector potential, and
no full equivalent of the gravitomagnetic field. Looking at the
effects of Earth's potential given by Eq.~(\ref{eq:87}), one
component of a satellite's acceleration is directed along the
potential gradient -- toward Earth's center. The other, containing
$\bf v$, is in the direction of the satellite's motion. For a
perfectly spherical Earth, that equation gives no accelerations
perpendicular to the orbital plane, or orbital precession due to
Earth's rotation.

A controversial analysis of the LAGEOS and LAGEOS II satellite
orbits by Ciufolini and Pavlis~\cite{cp} finds a Lens-Thirring
precession 99\% of general relativity's, with $\pm$10\% error. (The
effect is tiny compared to others such as those of Earth and ocean
tides, which must be estimated with great accuracy.) Iorio~\cite{li}
finds their calculations flawed and that no reliable measurement was
made. A definitive measurement of Lense-Thirring precession is
expected from NASA's Gravity Probe B, launched in April 2004. That
experiment reads the orientation of a sensitive gyroscope carried by
a satellite in polar orbit, with respect to a selected
star~\cite{cw}.

Gravity Probe B can be regarded as a test of Mach's principle -- a
pillar of Einstein's theory but not this. According to the former,
the gyroscope will rotate with its local inertial frame, and is
predicted to undergo a Lense-Thirring precession of .042
arcseconds/year. In this theory, since there are no rotating
inertial frames, a null measurement of the Lens-Thirring effect is
predicted. (The probe will also measure de Sitter precession,
where the gyroscope's direction within the satellite's orbital
plane changes~\cite{cw}. There the predictions of the two theories
are identical.)

\section{Gravitational Radiation}

The strongest evidence of gravitational waves at present is the
orbital decay of the binary pulsar, PSR 1913+16.  This star orbits
its companion with a short period of 7.75 hours, in a highly
eccentric orbit ($\epsilon$ = 0.617).  Since pulsars have the
regularity of atomic clocks, the orbital period and other parameters
of the system can be determined very accurately from pulse arrival
times. As shown by Weisberg and Taylor~\cite{wt}, the observed
orbital period shift agrees closely with the gravitational radiation
damping predicted by general relativity.

It's assumed the stars suffer no tidal effects and the surrounding
space is free of matter, so the loss of orbital energy is due
entirely to gravitational waves.  The loss is calculated from an
equation derived by Peters and Mathews~\cite{pm}.  That gives the
average power radiated by point masses in Keplerian orbits, based
on the weak-field, slow-motion approximation of Einstein's
gravitation. Acknowledging dissension on this point, Peters and
Mathews assume the energy carried by conventional gravitational
waves is real and positive, citing the analogy to
electromagnetism.

Several parameters needed for the calculation, the stellar masses
and projected axis of the pulsar orbit, are not directly available
from the pulsar data.  Taylor and Weisberg obtain the missing
parameters by solving three simultaneous equations, including one
for the periastron advance, and another for the combined
transverse Doppler and gravitational frequency shifts.  (The
equations for those effects are the same here.) The resulting
energy loss has been found to agree with the observed orbital
period shift to an accuracy of about 0.4\%~\cite{jht}.

Feynman~\cite{fls2} has pointed out that, even in classical
electromagnetism, there is ambiguity in the usual derivation of
radiated energy. The change in field energy for an infinitesimal
volume of space is described by
\begin{equation}
- \frac{\,\partial u}{\,\partial t} \:=\: \nabla \cdot {\bf S} \,+\,
{\bf E} \cdot {\bf j} \label{eq:99}
\end{equation}
where $u$ is the field energy density, ${\bf S}$ is the field
energy flow normal to the volume's surface, ${\bf E}$ the electric
field and ${\bf j}$ is the electric current density.

As done originally by Poynting, when $u$ and ${\bf S}$ are defined
as
\begin{equation}
u \:\equiv\: \frac{\,\epsilon_0}{\,2} \;{\bf E} \cdot {\bf E} \:+\,
\frac{\:\epsilon_0 c^2}{\,2}\;{\bf B} \cdot {\bf B} \label{eq:100}
\end{equation}
and
\begin{equation}
{\bf S} \:\equiv\: \epsilon_0 c^2 \,{\bf E}\times {\bf B}
\label{eq:101}
\end{equation}
Maxwell's equations can be used to show these expressions agree with
Eq.~(\ref{eq:99}).  However, Feynman notes that the latter is also
satisfied by other of combinations of $u$ and ${\bf S}$, infinite in
number, and we have no way of proving which is correct.
Eqs.~(\ref{eq:100}) and~(\ref{eq:101}) are believed correct because
they are the simplest pair of expressions in agreement with
experiment.

There is no ambiguity in Eq.~(\ref{eq:99}).  From energy
conservation, the work done by an electric field on a stream of
charged particles with constant velocity must be radiated
accordingly. (See Feynman~\cite{fls2}.) To arrive at the
gravitational counterpart of that equation, we'll consider the work
done on a stream of particles with constant velocity in a
gravitational field.

From Eq.~(\ref{eq:83}), the energy (Hamiltonian) for a particle in a
gravitational potential is
\begin{equation}
E \:=\:  \frac{m_{00} c_0^2}{\sqrt{1\,-\,v^2\!/c^2}} \:
e^{\Phi_{\!g}/c_0^2}
\end{equation}
For weak potentials and low velocities, that can be approximated
as
\begin{equation}
E \:\cong \: \frac {m_{00} c_0^2} {\sqrt{1\,-\,v^2\!/c^2}} \,+\,
m_{00} \Phi_{\!g}
\end{equation}
Then, for a single particle entering a gravitational potential at
a constant velocity, the energy lost to radiation is given by
\begin{equation}
\Delta E \:\cong \:  m_{00} \Phi_{\!g} \label{eq:104}
\end{equation}

In addition to the particle's energy, we'll need to account for
energy associated with its surrounding potential. Again, a
gravitational potential represents a region of diminished vacuum
energy. (From the extended principle of relativity, vacuum energy
diminishes along with that of ordinary particles. If not, a
measurement of the Casimir effect wouldn't give the same result in a
uniform gravitational potential.)

In electromagnetism, the potential due to a uniformly moving charge
doesn't change when it enters a different external potential.  But
there {\em is} a such a change for a particle entering a
gravitational potential. From Eq.~(\ref{eq:17}), its rest mass $m_0$
increases as $e^{-3\Phi_{\!g}/c_0^2}$. And from Eq.~(\ref{eq:6}),
its own potential strengthens proportionately. (Section 8 showed
this effect also influences the lunar orbit.)

The resulting loss of vacuum energy is likewise proportional to
the increase in the particle's rest mass.  Where $\Delta E_\Phi$
is the change in vacuum energy due to the particle's gravitational
potential, and the factor of proportionality between mass and
energy is the usual $c^2$, that be expressed as
\begin{equation}
\Delta E_\Phi \:=\:  -(\Delta m_{0}) c^2
\end{equation}
For weak potentials, the transformation of rest mass then gives
\begin{equation}
\Delta E_\Phi \:\cong\: 3 m_{00} \Phi_{\!g} \label{eq:106}
\end{equation}

According to the gravitational wave equation, Eq.~(\ref{eq:9}), as
a particle's rest mass increases, the change in its potential
propagates outward from the particle at the local speed of light.
Where the gravitational potential strengthens and vacuum energy
diminishes, the principle of energy conservation requires a
corresponding energy flow out of the region. And from the last
equation, the amount is three times the flow of energy lost by the
particle itself to radiation.

From the combined effects described in Eqs.~(\ref{eq:104})
and~(\ref{eq:106}), the radiation for a stream of particles with
constant velocity in a weak gravitational field is given by
\begin{equation}
- \frac{\,\partial u}{\,\partial t} \:\cong\: \nabla \cdot {\bf S}
\,-\, 4 \,\nabla \Phi_{\!g} \cdot {\bf j}_m
\end{equation}
where ${\bf j}_m$ is the mass current density.  The value here is
four times that suggested by the electromagnetic analogue,
Eq.~(\ref{eq:99}), when charge is replaced by rest mass in the
current density and the gradient of the potential is opposite in
sign. The factor of four difference arises because the equivalent of
charge is not conserved.

General relativity predicts no dipole gravitational radiation, and
Misner, Thorne and Wheeler~\cite{mtw3} note this would also be the
case for a gravity theory analogous to electromagnetism. There the
reason is that the center of mass of a gravitationally bound star
system doesn't accelerate. Consequently, the dipole contributions of
its individual stars cancel.

However, dipole radiation is possible in scalar-tensor gravity
theories, such as Brans-Dicke, which violate the strong equivalence
principle. In those, a body's acceleration depends partly on its
gravitational binding energy. If the stars in a binary system have
different binding energies, its inertial and gravitational centers
of mass will lie at slightly different points. And its gravitational
center will circulate as the stars orbit, resulting in dipole
radiation. Arzoumanian~\cite{za} finds that, for a pulsar-white
dwarf binary, where the binding energies differ markedly, the
effects of such dipole radiation may be observable. He doesn't find
that to be the case for PSR 1913+16, whose companion is another
neutron star.

Unlike the Brans-Dicke theory, this one effectively obeys the strong
equivalence principle -- at least for slow motion and weak
potentials. (As shown in Section 8, the solar accelerations of the
Moon and Earth are effectively identical, despite the difference in
their binding energies.) Hence measurable dipole radiation seems
unlikely, even for a pulsar-white dwarf pair.

Misner, Thorne and Wheeler~\cite{mtw3} note the quadrupole
gravitational radiation predicted by general relativity is four
times stronger than suggested by the analogy to electromagnetism.
Since the same factor of four difference exists in this theory,
there appears to be similar agreement with the gravitational
radiation damping observed in PSR 1913+16.

\section{A Different Cosmology}

Is the Riemann geometry of Einstein's general relativity necessarily
real?  Again, Poincar\'{e} held that nature singles out the simplest
of geometries, the Euclidean~\cite{ae4}.

A Riemann geometry like that conceived by Einstein can also be
used to describe ordinary optical systems~\cite{cl2}.  Similarly,
the speed of light is treated as absolute, while the ``optical
path distance" varies according to the refractive index. Although
it's possible to solve optics problems this way, of course
measurements with meter sticks show the true geometry of ordinary
optical systems is Euclidean.  Measurements of the distribution of
galaxies~\cite{ls} appear to be saying the same for the geometry
of the universe.

While the current geometric interpretation of general relativity
rests on an absolute speed of light in vacuo, that isn't the case
here.  Besides gravitation without space-time curvature, this
permits an alternate basis for the Hubble redshift: a decreasing
value of $c$.  Since the frequencies and wavelengths of de Broglie
waves depend on it, the wavelengths of atomic spectra would also be
shifted.

In accord with Eq.~(\ref{eq:10}), the diminishing speed of light is
taken to represent a gradual strengthening of the universe's overall
gravitational potential. (This effect may be attributable in part to
a transfer of energy between different scales in the
universe~\cite{bk}.) The extension of Poincar\'{e}'s relativity
principle to uniform gravitational potentials is also taken to hold
again. Consequently, while $c$ is evolving, the locally observed
value and laws of physics don't change.

From Eqs.~(\ref{eq:13}) and (\ref{eq:14}), the de Broglie
frequencies and wavelengths of particles diminish over time. I.e.,
the rates of clocks are slowing, and lengths of meter sticks are
contracting. For local observers, who see no change in their units
of length, the result is an apparent expansion of the universe.

Particle rest masses also increase, from Eq.~(\ref{eq:17}). That
brings a cosmological instability, not unlike the one in Einstein's
theory which once compelled him to introduce a cosmological
constant. While the increasing magnitude of the universe's overall
gravitational potential drives the masses of its particles higher,
that in turn drives the potential's magnitude up.

Suppose this cosmological potential has the value zero when light
leaves a distant galaxy, and $\Phi_{\!g}$ when it reaches Earth.
Then $c$ diminishes by the factor $e^{2 \Phi_{\!g}/c_0^2}$ over
that time. However, no difference arises in the relative speeds of
two successive wavefronts, so there is no shift in the light's
absolute wavelength. An observer on Earth sees an apparent
redshift, however, because the wavelength of a spectral reference
shortens by the factor $e^{\Phi_{\!g}/c_0^2}$. To the first order
in $\Phi_{\!g}/c_0^2$, the relative wavelength observed is
\begin{equation}
\lambda \:\cong\: {\lambda_0}\,(\,1- \Phi_{\!g}/c_0^2\,)
\end{equation}
where $\lambda_0$ is the wavelength of a local spectral reference.

For low redshifts, the empirical relationship is
\begin{equation}
\lambda \:\cong\: {\lambda_0}\,(\,1+H d/c\,)
\end{equation}
where $H$ is the Hubble constant and $d$ a galaxy's distance. From
measurements using the Hubble Space Telescope~\cite{jrm}, $H$ has
been estimated at 71 km/sec/Mpc. Equating the final terms of the
last two equations gives
\begin{equation}
H d/c \:=\: H t \:=\: -\Phi_{\!g}/c_0^2
\end{equation}
where $t$ is the time light travels from a galaxy to Earth. And
the resulting rate of change for the cosmological potential is
\begin{equation}
\frac{\,d}{\,dt}\! \left( \frac{\,\Phi_{\! g}}{\,c_0^2} \right)
\,=\: -H
\end{equation}

(In principle, this cosmological redshift could be measured in a
two-beam interferometer with arms of unequal lengths.  However, if
the interferometer's components are physically connected, the arms
will contract as the cosmological potential strengthens. And the
resulting displacement of its mirrors will cause a Doppler blueshift
which cancels the redshift.)

Because the slowing of light signals is about twice that of clocks,
an apparent time dilation is also seen in distant objects. Like the
Big Bang model, this agrees with the observed lifetimes of type Ia
supernovae, proportional to their redshifted wavelengths~\cite{agr}.
No attempt has been made yet to model the cosmic microwave
background. (Since this is not a steady-state cosmology, this will
involve inferring conditions in the remote past, and at distances
where individual objects are not resolved.)  However we know at
least that the sky between resolvable objects would be filled with
longer-wavelength radiation from more distant ones.

As indicated by Eq.~(\ref{eq:26}), here the speed of light near a
massive body is always positive in isotropic coordinates, and light
rays can always escape in the radial direction.  Hence the event
horizons and black holes predicted by current general relativity
don't occur (Einstein denied their existence), and there is no
problem of information loss.  (For supermassive bodies such as
galactic nuclei, a subsequent paper will show the fraction of
emitted light escaping is close to zero.)

A survey of galactic black hole candidates by Robertson~\cite{slr}
finds a strong resemblance to neutron stars in all cases (with the
exception of the galactic nucleus), and no evidence of event
horizons.  He also finds the fluctuating X-ray spectra of galactic
candidates agree well with a model where infalling matter meets a
hard surface, rather than falling through an event horizon.

This gravity theory bears some mathematical resemblance to the
exponential metric theory of Yimaz~\cite{hy1}, where likewise there
are no event horizons.  (The Yilmaz theory is also
renormalizable~\cite{hy2}.) That was the basis of an early quasar
model by Clapp~\cite{rec}, in which quasars as represented as
massive star-like objects.  Those avoid the collapse predicted by
the Schwarzschild metric, and would show high gravitational
redshifts. Cumulative observations by Arp~\cite{ha}, of quasars
associated with disturbed, low-redshift galaxies, appear to support
such a model.

\section{Conclusions}

John Wheeler has urged general relativity be ``battle tested"
against fundamentally different theories of gravity.  But it seems
many believe this is no longer necessary. Measurements of general
relativistic effects, such as gravitational bending of starlight,
are often cited as proof of space-time curvature. The implicit
assumption is that no alternative is possible. The theory described
here suggests this assumption is unjustified.

As in electromagnetism, we can attribute gravitation to the direct
influence of potentials on quantum-mechanical waves.  Unlike the
``standard" Big Bang model based on Einstein's theory, this one
agrees with the flat large-scale geometry of the universe observed
and permits stars with ages well above 13 Gyr.  And, unlike general
relativity, this theory is immediately compatible with quantum
mechanics. This calls into question the need for a curved
space-time, its great mathematical complexity, and many degrees of
freedom.

In the upcoming results from the satellite experiment Gravity Probe
B, general relativity calls for a Lense-Thirring precession of the
gyroscope, while a null effect is predicted here. Measurement of the
second-order solar deflection of light in the LATOR experiment could
also distinguish clearly between the two theories.

\section*{Acknowledgments}

I would like to thank Kenneth Nordtvedt for pointing out an
important mistake in the original paper, where an improper
comparison was made between the second-order gravitational
frequency shift in Einstein's theory and this one. Also James
Hartle, Thibault Damour, Hrvoje Nikolic, Alexander Silbergleit,
Hugo Loaiciga, Albert Overhauser, Stanley Robertson and
Jean-Pierre Vigier for their comments. I \nolinebreak thank
Michael Perryman for an update on planned orbiting
interferometers. And I am very grateful to John Wheeler for his
enthusiasm and encouragement.

\section*{Appendix A: Derivation of $\bf c$}

In accord with Einstein's view that the theory of gravity should
be derived from a set of general principles, here we'll derive
Eq.~(\ref{eq:10}) for the speed of light.  These are the
principles adopted:

\begin{quote}
1) Absolute space and time (as in the preferred-frame relativity
of Lorentz and Poincar\'{e}~\cite{jsb}).

2) Superposition of gravitational potentials.

3) Relativity also holds for uniform gravitational potentials;
i.e., the observed laws of physics remain unchanged.
\end{quote}

We'll use an Einstein-style {\em Gedanken} experiment:  Imagine a
clock enclosed in a spherical shell of matter, where there is a
uniform gravitational potential. The clock can be seen from
outside through a small window, and an observer with a second
clock sits at a distant location, where the potential due to the
shell is negligible. We want to know how the observed clock rates
compare.

The rate of any clock is determined by the de Broglie frequencies of
its constituent particles; hence, we can use the frequency of a
representative particle to gauge the relative rates.  From quantum
mechanics, the energy of a particle at rest in a {\em weak}
gravitational potential is described by
\begin{equation}
h \nu \:=\: m_{00} c_0^2 \,+\, m_{00} \Phi_{\!g}
\end{equation}
where $\nu$ is the frequency and $m_{00}$ refers to the mass when
both the velocity and gravitational potential are zero.  Dividing
the particle energy at the inner clock's position by that at the
outer clock (where $\Phi_{\!g}$ vanishes) we we find the relative
rates are given by
\begin{equation}
\frac{\,\nu}{\,\nu_0} \:=\:  1 + \frac{\,\Phi_{\!g}}{\,c_0^2}
\end{equation}
when the gravitational potential is weak.

Next, we enclose everything in a second mass shell, positioning
another observer and clock outside that, where {\em its} potential
is vanishing. We also insure that the contribution of the second
shell to the cumulative potential is equal.  By our third
principle, the frequency ratio measured by the inner observer is
unchanged, and those measured by the two observers are identical.
Numbering our clocks from the outside, the subscript $0$ will
indicate the outermost, and $2$ the inner one. We have
\begin{equation}
\frac{\,\nu_1}{\,\nu_0} \:=\:  \frac{\,\nu_2}{\,\nu_1} \:=\: 1 +
\frac{\,1}{\,2}\frac{\,\Phi_{\!g}}{\,c_0^2}
\end{equation}
where $\Phi_{\!g}$ represents the cumulative potential from both
shells. Hence the frequency ratio for the innermost vs. the
outermost clock is
\begin{equation}
\frac{\,\nu_2}{\,\nu_0} \:=\:  \left(1 + \frac{\,1}{\,2}
\frac{\,\Phi_{\!g}}{\,c_0^2}\right)^2
\end{equation}
The process can be repeated indefinitely, and for $n$ shells, we
get
\begin{equation}
\frac{\,\nu_n}{\,\nu_0} \:=\:  \left(1 + \frac{\,1}{\,n}
\frac{\,\Phi_{\!g}}{\,c_0^2}\right)^n
\end{equation}

The effect of a {\em strong} gravitational potential can be
represented as that of an infinite series of mass shells, each
producing a weak contribution to the total potential.  Taking the
limit as $n$ goes to infinity, we find
\begin{equation}
\frac{\,\nu}{\,\nu_0} \:=\: \lim_{n\to\infty} \left(1 +
\frac{\,1}{\,n} \frac{\,\Phi_{\!g}}{\,c_0^2}\right)^n \:=\;
e^{\Phi_{\!g} / c_0^2}
\end{equation}

As shown by the Collela-Overhauser-Werner experiment, de Broglie
wavelengths also diminish in a weak gravitational potential as
\begin{equation}
\frac{\,\lambda}{\,\lambda_0} \:=\:  1 +
\frac{\,\Phi_{\!g}}{\,c_0^2}
\end{equation}
For a strong potential, a similar {\em Gedanken} experiment
measuring wavelengths gives
\begin{equation}
\frac{\,\lambda}{\,\lambda_0} \;=\: e^{\Phi_{\!g} / c_0^2}
\end{equation}

The general velocity $V$ of quantum-mechanical waves is given by $V
= \lambda \,\nu$.  And from the extended principle of relativity,
$c$ and $V$ change identically. Hence
\begin{equation}
\frac{\,c}{\,c_0} \:=\: \frac{\,V}{\,V_0} \:=\; e^{2 \Phi_{\!g} /
c_0^2}
\end{equation}
Einstein: ``When the answer is simple, God is talking."

\section*{Appendix B: Equations of Motion}
Here we'll illustrate the use of the Lagrangian introduced in
Section 9, by deriving the equations of motion for a body in a
central gravitational field.  And we'll show they agree with the
equations for Mercury's orbit found earlier by the de Broglie wave
method.

Again, the inertial frame will be that where the Sun or central body
is at rest. Putting Eq.~(\ref{eq:81}) in terms of $\mu$ and polar
coordinates, the Lagrangian becomes
\begin{equation}
L \:=\: -\,E_{00}\: e^{-\mu/r} \sqrt{1-\,e^{4\mu/r}\!\left({\dot
r}^{\,2}+ r^2 \dot \theta^{\,2}\right)/\:c_0^2 }
\end{equation}
where the dots indicate time derivatives. Taking $\theta$ as the
generalized coordinate, the Euler-Lagrange equation of motion is
\begin{equation}
\frac {\,\partial L}{ \partial \,\theta} \,-\, \frac{\: d}{\, d t}
\left ( \frac {\,\partial L}{\, \partial \,\dot \theta}\, \right)
=\: 0 \label{eq:121}
\end{equation}

For $\partial L/ \partial \dot \theta$, we obtain
\begin{equation}
\frac {\,\partial L}{\, \partial \,\dot \theta} \:=\: \frac {r^2
\,\dot \theta \,E_{00} \:e^{3\mu/r} } {c_0^2
\,\sqrt{1-\,e^{4\mu/r}\!\left({\dot r}^{\,2}+ r^2 \dot
\theta^{\,2}\right)/\:c_0^2}} \label{eq:122}
\end{equation}
The square root in the denominator represents the quantity
$\sqrt{\,1-v^2/c^2}\,$, and from Eqs.~(\ref{eq:29})
and~(\ref{eq:32})), that can be expressed as
\begin{equation}
\sqrt{1-\,e^{4\mu/r}\!\left({\dot r}^{\,2}+ r^2 \dot
\theta^{\,2}\right)/\:c_0^2 } \:=\: \frac{E_{00} \,e^{-\mu/r}}{E}
\label{eq:123}
\end{equation}
Since the term $\partial L /\partial \theta$ in Eq.~(\ref{eq:121})
is zero, we also have
\begin{equation}
\frac{\, d}{\, d t} \left ( \frac {\,\partial L}{\, \partial
\,\dot \theta}\, \right) \,=\: 0
\end{equation}
From the last two equations, Eq.~(\ref{eq:122}) can be written as
\begin{equation}
\frac {\,\partial L}{\, \partial \,\dot \theta} \:=\: \frac {r^2
\,\dot \theta \,E \:e^{4\mu/r}} {c_0^2} \:=\: C
\end{equation}
where $C$ is a constant.

Again, from the principle of energy conservation, $E$ is a
constant for a freely orbiting body.  And dividing by $E/c_0$
gives
\begin{equation}
\frac {\,r^2 \,\dot \theta \:e^{4\mu/r}} {\,c_0} \:=\: k
\end{equation}
where $k$ represents another constant.  This is our earlier
Eq.~(\ref{eq:37}), in polar coordinates. Rearranging, the time
derivative of $\theta$ is
\begin{equation}
\dot \theta \:=\: \frac {\,k \,c_0 \,e^{-4\mu/r}} {\,r^2}
\end{equation}
Putting this expression for $\,\dot \theta \,$ into
Eq.~(\ref{eq:123}) and solving for $\, \dot r$, we also obtain
\begin{equation} \dot r \:=\: \frac {\,c_0 \,e^{-4\mu/r} \sqrt{\,r^2
\left( e^{4\mu/r} - \frac {E_{00}^2} {E^2} \,e^{2\mu/r} \right)
-k^2}} {r}
\end{equation}

Taking the ratio $\dot r/ \dot \theta$ eliminates their time
dependence, and gives the first-order differential equation of the
orbit:
\begin{equation}
\frac {\,dr} {\,d \theta} \:=\: \frac {r \sqrt{\,r^2 \left(
e^{4\mu/r} - \frac {E_{00}^2} {E^2} \,e^{2\mu/r} \right) -k^2}} {k}
\label{eq:129}
\end{equation}
The second-order differential equation of the orbit can be obtained
by differentiating this one.  After substituting for square root
terms via Eq.~(\ref{eq:129}), the result can be arranged as
\begin{equation}
\frac {\,d^2 r} {\,d \theta ^2} \,=\, \frac {1}{r} \!\left(\frac
{dr} {d \theta} \right)^{\!2} \!+\, \frac{r^3 \!\left( e^{4\mu/r}
\!- \frac {E_{00}^2}{E^2} e^{2\mu/r} \right) \!- \mu r^2
e^{4\mu/r} \!- \mu r^2 \!\left( e^{4\mu/r} \!- \frac
{E_{00}^2}{E^2} e^{2\mu/r} \right)}{k^2}
\end{equation}
Also, from Eq.~(\ref{eq:129}),
\begin{equation}
r^2 \!\left( e^{4\mu/r} - \,\frac {\,E_{00}^2}{\,E^2} \,e^{2\mu/r}
\right) \,=\: \frac{\,k^2}{\,r^2} \left(\frac {\,dr} {\,d \theta}
\right)^{\!2} \!+\, k^2
\end{equation}
After this substitution, the previous equation reduces to
\begin{equation}
\frac {\,d^2 r} {\,d \theta ^2} \:=\: \frac {\,2 - \mu/r} {r}
\left(\frac {\,dr} {\,d \theta} \right)^{\!2} \!+\, r \,-\, \mu
\,-\, \frac {\,r^2 \mu \,e^{4\mu/r}} {k^2}
\end{equation}
which can be compared directly to the Newtonian case.

Differentiating both sides of Eq.~(\ref{eq:35}) to remove the
integral and rearranging, the result is Eq.~(\ref{eq:129}).  This
shows the solution of these differential equations is the same
obtained by the de Broglie wave method, which is in agreement with
Mercury's orbit. And we have the possibility that matter waves are
the basis of {\em all} mechanics.

\vfill\eject

\end{document}